\documentclass{article}
\usepackage{arxiv}

\usepackage{amsmath}
\usepackage{mathrsfs}
\usepackage{graphicx}
\usepackage[hidelinks]{hyperref}
\usepackage{amsmath}
\usepackage{algorithm,algpseudocode}
\usepackage{subcaption}
\usepackage{multirow}
\usepackage[numbers]{natbib}

\usepackage[title]{appendix}

\newcommand{\figref}[1]{Figure \ref{#1}}
\newcommand{\appfigref}[1]{Appendix Figure \ref{#1}}
\newcommand{\apptabref}[1]{Appendix Table \ref{#1}}
\newcommand{\Secref}[1]{Section \ref{#1}} 
\newcommand{\ud}{\mathop{}\!\mathrm{d}}
\renewcommand{\algref}[1]{Algorithm \ref{#1}}

\title{Incorporating additional evidence as prior information to resolve non-identifiability in Bayesian disease model calibration. A tutorial.}

\author{
  Daria~Semochkina\\
  School of Mathematical Sciences\\
  University of Southampton\\
  \texttt{d.semochkina@soton.ac.uk} \\
   \And
 Cathal D.~Walsh\\
  TCD Biostatistics Unit, Public Health and Primary Care, School of Medicine.\\
  Trinity College Dublin\\
  \texttt{Cathal.Walsh@tcd.ie} \\
}

\begin{document}
\maketitle

\begin{abstract}
Disease models are used to examine the likely impact of therapies, interventions and public policy changes. Ensuring that these are well calibrated on the basis of available data and that the uncertainty in their projections is properly quantified is an important part of the process. The question of non-identifiability poses a challenge to disease model calibration where multiple parameter sets generate identical model outputs. 

For statisticians evaluating the impact of policy interventions such as screening or vaccination, this is a critical issue. This study explores the use of the Bayesian framework to provide a natural way to calibrate models and address non-identifiability in a probabilistic fashion in the context of disease modelling. We present Bayesian approaches for incorporating expert knowledge and external data to ensure that appropriately informative priors are specified on the joint parameter space.
These approaches are applied to two common disease models: a basic Susceptible-Infected-Susceptible (SIS) model and a much more complex agent-based model which has previously been used to address public policy questions in HPV and cervical cancer. The conditions which allow the problem of non-identifiability to be resolved are demonstrated for the SIS model.  For the larger HPV model an overview of the findings is presented, but of key importance is a discussion on how the non-identifiability impacts the calibration process. Through case studies, we demonstrate how informative priors can help resolve non-identifiability and improve model inference. We also discuss how sensitivity analysis can be used to assess the impact of prior specifications on model results. Overall, this work provides an important tutorial for researchers interested in applying Bayesian methods to calibrate models and handle non-identifiability in disease models. 
\end{abstract}


\maketitle

\section{Introduction}

Disease modelling plays a crucial role in epidemiology by providing insights into disease transmission dynamics and informing public health interventions. Mathematical models capture these dynamics through a set of parameters, such as rates related to transmission, recovery, and mortality. Such parameters quantify key features of the disease. Estimating those parameters accurately is essential for reliable model predictions. The process of estimating model parameters is termed calibration. This work was motivated by the authors' previous experience dealing with challenges which arose while calibrating disease models for the purposes of informing policy on screening programmes\cite{whyte2011bayesian}, vaccination strategies\cite{usher2008cost} and while involved in modelling during the COVID pandemic\cite{duggan2024age}.

Bayesian inference offers a powerful framework for parameter estimation in disease models. It is often desirable that the uncertainty associated with any forecasts is quantified in the distribution of the parameters obtained through calibration. The Bayesian framework is a natural one to use for solving inverse problems such as model calibration. The goal of Bayesian inference is to sample from a probability distribution which is most compatible with a combination of data and a prior distribution describing the parameter space. The distribution of interest is often high-dimensional and does not have a standard form. Markov chain Monte Carlo (MCMC) methods are powerful tools for solving inverse problems and obtaining a sample from the posterior distribution of interest\cite{whyte2011bayesian}. We examine the Bayesian approach to calibration, how the stochastic samples from the parameter sets are summarised and how to deal with a key challenging aspect, namely that of non-identifiability.

A model is said to be \emph{non-identifiable} if two or more distinct sets of parameter values can produce the same observed data. That is, we cannot uniquely determine the true parameter values from the available data. For example, in a simple linear regression model, if two predictor variables are perfectly correlated, it becomes impossible to isolate the individual effects of each variable on the response. The issue of identifiability can occur in basic mixture models (such as a simple two-component mixture) and has been seen by the current authors in other situations, such as that of label switching in latent variable models \cite{walsh2006latent}. This can lead to difficulty in interpreting posterior distributions and drawing reliable inferences about the model parameters. In the context of disease modelling, non-identifiability can occur for various reasons, such as limitations in the available data or inherent structural features of the model\cite{alarid2018nonidentifiability,gustafson2015bayesian,wang2013identifiability,arendt2012quantification,arendt2012improving}.

This study addresses the problem of non-identifiability in Bayesian disease model calibration. We explore the use of informative prior information, derived from expert knowledge or external data, to resolve non-identifiability issues. We focus on two common disease models: the Susceptible-Infected-Susceptible (SIS) model and the agent-based Susceptible-Infected-Recovered-Deceased (SIRD) HPV model with probabilistic transition rates. We will demonstrate how informative priors can help to improve the precision of posterior estimates. Additionally, we discuss the importance of sensitivity analysis to evaluate the robustness of model results to variations in prior specifications.

\section{Disease Models and Bayesian Model Calibration Basics}\label{calib:M-H}

This section introduces the fundamental concepts of disease modelling and the Bayesian approach to model calibration. We explore how Bayesian inference provides a powerful framework for parameter estimation in disease models. We outline the challenges that arise with non-identifiability in model calibration and common causes of non-identifiability. Several recommendations are provided to tackle non-identifiability.

\subsection{Compartmental Models for Disease Dynamics}

Disease models often use a compartmental approach to describe the progression of infections within a population. These models categorise individuals into compartments based on their disease status and define transitions between these states through mathematical equations. This framework has been instrumental in understanding the dynamics of disease transmission, evaluating control measures, and guiding public health policy\cite{haeussler2018dynamic}.

One of the earliest examples of compartmental models for disease dynamics is the SIR model, introduced by Kermack and McKendrick\cite{kermack1927contribution} in 1927, which includes states of susceptible (S) and infectious (I) as well as an additional compartment: recovered (R). In the SIR model, individuals transition from susceptible to infected and then to recovered, where they gain immunity. The SIR model underpins many critical epidemiological concepts, such as the basic reproduction number, $R_0$, and the herd immunity threshold. This model is widely used for studying outbreaks of infectious diseases like measles and influenza, where immunity plays a significant role in the population's response to an epidemic.

Another foundational example is the SIS model\cite{hethcote1976qualitative}, where individuals are grouped into two compartments: susceptible (S) and infected (I) (see \figref{fig:A_B_diagram}). Susceptible individuals can contract the disease through contact with infected individuals at a rate defined by the transmission parameter $\beta$. Infected individuals recover at a rate $\gamma$ but lose immunity and return to the susceptible state. This structure is ideal for diseases where immunity is not long-lasting, such as certain bacterial infections.

For diseases with an incubation period, the SEIR model introduces an exposed (E) compartment, representing individuals who are infected but not yet infectious\cite{anderson1991infectious}. This extension captures the dynamics of diseases like COVID-19\cite{he2020seir,carcione2020simulation,lopez2021modified} or influenza\cite{arino2008model,colizza2007modeling}. Individuals move from the susceptible state to the exposed state upon infection at a rate 
$\beta$, then progress to the infected state after an incubation period at rate $\sigma$, and finally recover at rate $\gamma$.

The work of Kermack and McKendrick also extended to include endemic diseases by incorporating demographic factors such as births, deaths, and partial immunity\cite{kermack1932contributions,kermack1933contributions}. These models provide a framework for analysing diseases that persist in populations over long periods, such as malaria or tuberculosis, and are capable of capturing the effects of interventions like vaccination.

Compartmental models remain indispensable for understanding disease dynamics, predicting epidemic trajectories, and guiding public health interventions. They have been applied in diverse contexts, from classical studies of measles outbreaks to the global response to the COVID-19 pandemic. For example, the SEIR model was crucial in forecasting the impact of social distancing and vaccination strategies during the early stages of COVID-19\cite{xu2021control,mcquade2021control}. Similarly, the foundational SIR model continues to inform vaccine allocation and predict outbreak sizes. These applications highlight the flexibility and enduring relevance of compartmental models in addressing emerging and existing infectious diseases

\subsection{Parameter Estimation and Key Concepts in Bayesian Inference}

Estimating parameters (e.g., transmission rate, recovery rate) of a disease model allows us to understand and make predictions about the disease dynamics. Bayesian inference helps us quantify our knowledge about these parameters in terms of a joint probability distribution. In general, Bayesian inference provides a statistical framework for reasoning under uncertainty. It has the following key constructs:

\begin{itemize}
    \item \emph{Prior Distribution}:  Represents our beliefs or knowledge about the parameter values without consideration of directly observed data. Priors can be informative (based on expert knowledge or previous studies) or non-informative (weak priors with minimal influence). Information in the prior is separate from that obtained during a study or experiment used for model fitting or calibration.
    
    \item \emph{Likelihood Function}:  The likelihood is the function through which new data inform parameter estimates. It quantifies the probability of observing the data given specific parameter values. The higher the likelihood, the better the data fits the model with those parameter values.
    \item \emph{Posterior Distribution}:  Combines the prior information with the information from the newly observed data. It represents the updated beliefs about the parameter values after considering both the prior and the observed data.  The posterior distribution is fully specified by the combination of prior and likelihood.
\end{itemize}

In the context of disease modelling the prior distribution could represent expert knowledge about the typical range of values taken by the transmission rate of a disease. The likelihood function would quantify how well a specific transmission rate value fits the observed disease prevalence data. The posterior distribution would give us a more precise estimate of the transmission rate, taking into account both prior knowledge and the data.

\subsection{Implementation of Bayesian Model Calibration}

There are various software implementations and computational methods for implementing Bayesian model calibration \cite{kennedy2001bayesian,gardner2019sequential} with our implementation here using Markov Chain Monte Carlo (MCMC) methods\cite{geyer1992practical,gilks1995markov,tarantola2005inverse,robert2010introducing}\cite{speich2021sequential,whyte2011bayesian, bliznyuk2008bayesian}. These methods involve generating a large number of sets of parameter values from the posterior distribution which can be used to draw inferences about the model parameters.

The Metropolis-Hastings algorithm (M-H)\cite{metropolis1953equation,hastings1970monte} is an MCMC algorithm for obtaining a sample from a probability distribution $f$, for which direct sampling is difficult or impossible. The M-H algorithm enables sampling from complex distributions to obtain (almost) independent samples directly from such distributions\cite{robert2004metropolis,tarantola2005inverse}. The algorithm is based on proposing values, sampled from some \emph{proposal distribution}. These values are then accepted or rejected with a certain \emph{probability of acceptance}, based on how likely those values are for some distribution of interest $f$, generating a Markov chain of parameter values that eventually converges to the target distribution. See Appendix \ref{app:M-H} for details. 

The steps in the Metropolis-Hastings Algorithm are outlined as follows. Step 1 (Initialisation): Choose a starting value for the parameter vector $\boldsymbol{x}$. 
Step 2 (Proposal Generation): Propose a new parameter vector based on a proposal distribution (e.g., a random walk, a multivariate normal distribution, etc.).
Step 3 (Acceptance Probability): Calculate the acceptance probability based on the ratio of the posterior densities at the proposed and current parameter values. Note that because the algorithm uses a ratio, a normalising constant is not required\cite{hastings1970monte}, which is the usual obstacle in acquiring the posterior distribution.
Step 4 (Acceptance or Rejection): Randomly decide whether to accept or reject the proposed value based on the acceptance probability. If accepted, the proposed value becomes the new state of the chain; otherwise, the current state is retained.
Iteration: Repeat steps 2-4 for a specified number of iterations.

Some practical considerations are worth reviewing here. 

\begin{itemize}
    \item Firstly, the choice of proposal distribution can significantly impact the efficiency of the algorithm. A well-tuned proposal can improve mixing and reduce autocorrelation of the generated sample.
    \item Secondly, it is essential to assess whether the MCMC chain has converged to the target distribution. Visual diagnostics of trace plots and tools like the Gelman-Rubin statistic can be used for this purpose\cite{gelman1992inference}.
    \item Thirdly, the initial samples from the MCMC chain may not be representative of the target distribution. A burn-in period is often used to discard these early samples.
    \item Finally, to reduce the autocorrelation of the samples, the chain can be thinned by selecting every $n$-th point.
\end{itemize}
Several statistical software packages (e.g., R\cite{R}, Stan\cite{carpenter2017stan}, WinBUGS\cite{lunn2000winbugs}) provide implementations of the Metropolis-Hastings algorithm and other MCMC methods. Regardless of the software used, the considerations about non-identifiability arise, and the solution using our proposed approach applies. In this tutorial, we will focus on implementing the Metropolis-Hastings algorithm in particular and addressing these practical considerations. We will demonstrate that by doing this, researchers can obtain reliable parameter estimates for their disease models, even in the presence of non-identifiability.

\subsection{Challenges}\label{Limitations}

A number of challenges exist when calibrating models using limited data. These include difficulties related to the high dimensionality of the parameter space, multi-modality of the likelihood function, and, more importantly, the question of identifiability for model calibration. \emph{High-dimensional parameter spaces}: When models involve numerous parameters, the computational burden of exploring the posterior distribution can be substantial. \emph{Multi-modality}: The likelihood function may have multiple local maxima, leading to difficulties in convergence. \emph{Computational intensity of models}: Models like micro-simulation or agent-based models can be computationally expensive to simulate and it is necessary to ensure the sampling algorithm is efficient.

The challenge we will focus our attention on in this tutorial is that of identifiability\cite{alarid2018nonidentifiability}. As mentioned before, non-identifiability arises where the data are consistent with at least two (often a continuum of) different sets of parameter values. For example in the case of disease progression, the data could be consistent with situations where the rate of progression is fast and the rate of recovery is also fast but equally, it could be the situation where the rate of recovery is slow with a slow rate of progression. In this situation, commonly collected data such as that on the prevalence of a disease gives only the total number of cases of this disease in a given population at a specific time. This data will be consistent with individuals progressing and recovering quickly as well as being consistent with individuals both progressing and recovering slowly. No information in the prevalence data can distinguish between those scenarios. In this situation, a good solution to overcoming the question of identifiability of the progression rate would be to elicit prior information about the recovery rate either from experts or through direct observation of cases.

\subsection{Non-Identifiability in Disease Models}\label{Non-Identifiability}

Non-identifiability arises when multiple parameter sets can generate the same model output. In the context of disease models, different model parameter combinations can produce indistinguishable disease dynamics patterns. There are several examples of non-identifiability. \emph{Simultaneous changes in incidence and recovery rates} (i.e.~increasing both the incidence rate and the recovery rate) can lead to similar epidemic curves, making it difficult to distinguish between these two effects. Models with \emph{hidden states}, such as latent infections or asymptomatic cases, can introduce non-identifiability if the data do not directly observe these states. \emph{Structural non-identifiability} occurs when the model structure itself is inherently non-identifiable, regardless of the data. For example, certain compartmental models may have symmetries that lead to non-identifiability.

There are some major implications of non-identifiability\cite{alarid2018nonidentifiability}. 
Non-identifiability can lead to wide credible intervals or flat posterior distributions, making it difficult to draw precise inferences about model parameters. If non-identifiability is not addressed, it can lead to incorrect model predictions and misleading conclusions. Non-identifiable models can be difficult to calibrate using standard methods, as the likelihood surface may have multiple modes or flat regions.

Several methods can be used to detect non-identifiability. 
\begin{itemize}
    \item \emph{Profiling the likelihood} function over different parameter values could help identify regions of the parameter space where the likelihood is relatively flat.
    \item The \emph{Fisher information matrix} can be used to assess the curvature of the likelihood surface and identify regions of non-identifiability. However, this is not an option for non-parametric models which are the most used in disease modelling.
    \item Introducing \emph{additional latent variables} can sometimes help to resolve non-identifiability issues.
    \item It was also demonstrated that using \emph{multiple observed responses} that all depend on some common set of calibration parameters\cite{arendt2012improving} could in some cases improve identifiability.
\end{itemize}

Several approaches can be applied to addressing non-identifiability. Incorporating \emph{strong prior information} can help to break the degeneracy and improve identifiability. In some cases, it may be possible to \emph{reformulate the model} to address non-identifiability. For example, introducing additional constraints or assumptions can help to narrow down the parameter space. Adding \emph{additional data} or measurements can sometimes help to resolve non-identifiability. By understanding the challenges posed by non-identifiability and employing appropriate methods, researchers can improve the reliability and interpretation of their disease models. 

An alternative approach to addressing non-identifiability is to reframe the problem by focusing on identifiable parameters or combinations of parameters. For instance, Gustafson \cite{gustafson2015bayesian} outlines methods for partially identified models, where inference is restricted to identifiable quantities. Additionally, marginalising over non-identifiable parameters within a Bayesian framework can still provide useful insights about identifiable components. These strategies complement the use of informative priors and may be particularly useful in cases where eliciting prior information is challenging.

To address the challenges of non-identifiability, this tutorial focuses on incorporating informative prior distributions for model parameters. This approach is supported by the work of Neath and Samaniego \cite{neath1997efficacy}, who demonstrated the effectiveness of prior information in handling a non-identifiable model. They have demonstrated that the extent of the improvement that a posterior estimate provides over a prior estimate is potentially quite large. However, the potential for the posterior estimate of a non-identifiable parameter being inferior to the prior because of Bayesian updating is bounded. By specifying appropriate priors, we can constrain the parameter space and improve the precision of our posterior estimates.

\section{Incorporating Prior Information to Address Non-Identifiability}\label{sec:approach}

In practice, non-identifiability can be diagnosed when there is a lack of convergence of chains produced by the Metropolis-Hastings algorithm. For example, when the posterior distributions of parameters are flat and broad, or when there is cross-correlation between sampled values. 

To assess the convergence of MCMC chains, visual diagnostics of trace plots such as \figref{fig:SIS chains} and \figref{fig:HPV_no_prior_vs_prior} are crucial. Trace plots provide valuable insights into the chains' behaviour. They help visualise the evolution of parameter values over iterations, with convergence indicated by a stable, random walk. A well-converged chain will exhibit a stable, random walk with no discernible trends (the right-hand side of both plots). Conversely, trends, cycles, or abrupt jumps indicate non-convergence (the left-hand side of both plots). By examining these plots, we can identify potential issues like non-convergence or poor mixing and take corrective actions.

To resolve non-identifiability, the suggested approach is to include information about some of the variables where this is obtainable. This can be done by carefully considering whether additional information has been collected or is available from other sources. For example, duration of illness may not be available locally, but a prior may be obtained by considering international evidence that is emerging. Note that from a structural and inferential viewpoint, the prior contains information about the parameters that is not available in the likelihood.  It does not necessarily have to have arisen or been obtained before the data which is used in the likelihood function. As in the examples above, observing cases directly and noting the time to recovery will give us different information than prevalence estimates when trying to ascertain rates of incidence or occurrence of disease.

In algorithmic terms, the process of calibration would proceed as follows:
\begin{itemize}
    \item Identify model outputs of interest (targets) and populate these with collected data.
    \item Specify priors for parameters with a `first pass', perhaps reasonably vague specification.
    \item Run the MCMC calibration process.
    \item Identify parameters that are unidentifiable, together with those that are correlated.
    \item Consider whether additional evidence may be obtained from external sources.
    \item Quantify information about this in terms of a `prior' probability distribution.
    \item Rerun MCMC sampling to calibrate the model conditional on this additional evidence.
\end{itemize}
By effectively incorporating prior information, we can improve the precision and reliability of parameter estimates in SIS models, especially when dealing with non-identifiability issues. 

\section{Example: SIS model}\label{sec:example:SIS}

\begin{figure}[!h]{}
 \centering
 \begin{subfigure}{.5\textwidth}
 \begin{center}
 \includegraphics[width=.8\textwidth]{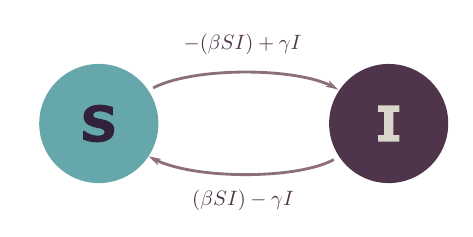}
 \end{center}
 \caption{}
 \label{fig:A_B_diagram}
 \end{subfigure}%
 \begin{subfigure}{.5\textwidth}
 \begin{center}
 \includegraphics[width=.8\textwidth]{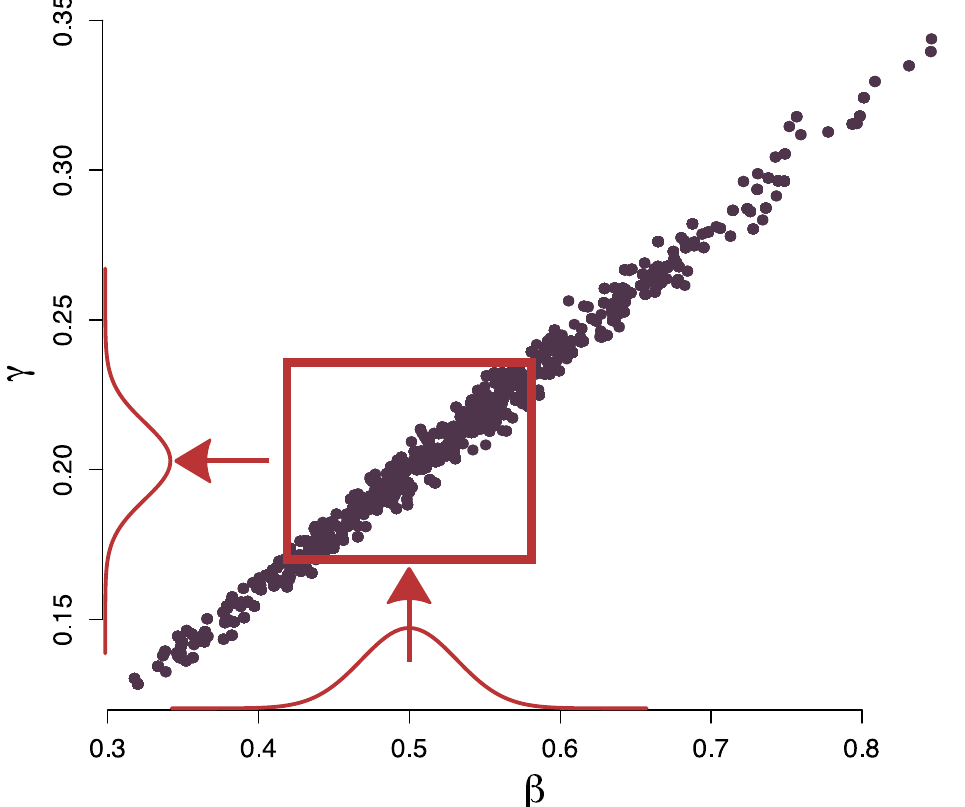}
 \end{center}
 \caption{}
 \label{fig:S_I_res}
 \end{subfigure}
 
 \begin{subfigure}{\textwidth}
 \begin{center}
 \includegraphics[width=.7\textwidth]{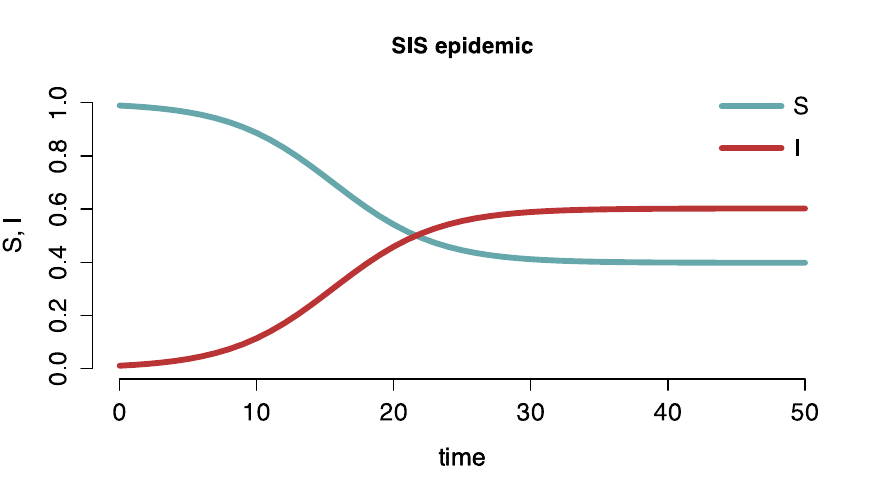}
 \end{center}
 \caption{}
 \label{fig:S_I}
 \end{subfigure}
 \caption{Simple example of an SIS model (a). Panel (b) shows the correlation between two parameters and the effect of a prior for one parameter ($\beta$) on the joint posterior space and marginal posterior distribution of $\gamma$. Panel (c) presents disease dynamics over time.}
 \label{fig:A_B}
 \end{figure}

To illustrate the challenges of non-identifiability further we provide an example in line with the model introduced in \figref{fig:A_B_diagram}. We consider a simple susceptible-infected-susceptible (SIS) compartmental model. This simple SIS compartmental model represents the dynamics of diseases with no permanent immunity. In this model, individuals can move between the susceptible (S) and infected (I) compartments. The transitions between the two states are described by
\begin{align*}
\ud S &= - (\beta\cdot S \cdot I) + \gamma \cdot I\\
\ud I &= (\beta\cdot S \cdot I) - \gamma \cdot I,
\end{align*}
where $\beta=c\cdot p$ and $\gamma=1/d$ with $c$ being the contact rate, $p$ the transition probability and $d$ the infectious period. Most of the population starts off in the susceptible state with $S=0.99$ and $I=0.01$. We treat this situation as a population where there has initially been disease introduced, and at some later stage, a stable situation has been observed where 15,000 of 25,000 individuals are identified as infected. The likelihood may be summarised as probabilities of being susceptible or infected (see \figref{fig:S_I} for an example of disease dynamic). After a period of active disease dynamics, the probabilities are stabilised around the 0.4 and 0.6 values for infected and susceptible compartments respectively. From the data collection point (i.e.~observing 15,000 infected individuals in a population of 25,000 with some uncertainty added for observational error), this could be transformed into having normal distributions corresponding to probabilities of being in each compartment introduced as targets. These targets ($N(0.4, 0.01)$ and $N(0.6, 0.01)$) are then used in the  Metropolis-Hastings algorithm. In particular, this means that the full SIS model is run for every proposed new input parameter combination and the produced output is compared to the target distributions. The likelihood is calculated as the negative log of the product of the two probability density functions of the two values output by the model normalised by the values of those density functions at the modes.
Implicit in this is that a uniform prior is used for $p$ and $c$.

The SIS model can exhibit non-identifiability, particularly when the data are limited or noisy. For example, it can be difficult to distinguish between high transmission rates with high recovery rates and low transmission rates with low recovery rates if the overall prevalence of the disease remains relatively constant as described in \Secref{Limitations}. To address non-identifiability in models, prior information can be invaluable. This information can come from various sources, including:
\begin{itemize}
    \item \emph{Expert knowledge.} Epidemiologists or infectious disease experts may have prior beliefs about the transmission rate ($\beta$) based on their understanding of the disease, its biology, and previous outbreaks.
    \item \emph{Historical data.} Data from past outbreaks of similar diseases can provide insights into the likely range of transmission rates.
    \item \emph{Biological constraints.} Biological factors, such as the average contact rate or the duration of infectiousness, can be used to inform the prior distribution for $\beta$.
\end{itemize}
To incorporate prior information into the SIS model, we can specify a prior distribution for the transmission rate ($\beta$). For example, if expert opinion suggests that $\beta$ is likely to be between 0.4 and 0.6, we could use a uniform distribution over this range. 

It is essential to assess the impact of the prior distribution on the posterior estimate of $\beta$. Sensitivity analysis can be used to evaluate how the posterior varies with different prior specifications. If the posterior is highly sensitive to the prior, it suggests that the data are not very informative and that the prior is playing a dominant role in shaping the estimate. That means that without a strong informative prior it will be impossible to have a focused posterior and therefore make any meaningful inference.

After running a simple Metropolis-Hastings algorithm for $c,\ p$ and $d$ with a uniform prior and the two normal distributions ($N(0.4, 0.01)$ and $N(0.6, 0.01)$) as our targets we plot the resulting $\beta$ and $\gamma$ values in \figref{fig:S_I_res}. It is noticeable that those are almost perfectly correlated. In panel (b) of the figure, the impact of specifying a prior for one variable is shown. The density in panel (b) represents an informative prior, with arrows illustrating how it impacts upon the joint posterior and hence the marginal posterior of $\gamma$. The annotated density on the horizontal axis for $\beta$ illustrates how the information in that parameter restricts the joint posterior, and thus provides indirect information on $\gamma$, the marginal distribution of which is now affected. Even though no information about the second variable is added, conditioning on a narrower prior distribution for $\beta$ restricts the joint posterior space to a narrower range, thus impacting the marginal posterior distribution for $\gamma$ which is induced due to the correlation between the parameters. Additionally, because $\beta=c\cdot p$, $c$ and $p$ are perfectly correlated. This leads to the chains for  $c,\ p$ and $d$ not converging due to the complex links between their values and the observational equivalence of the resulting targets.
  \begin{figure}[!h]{}
 \centering
 \includegraphics[width=.8\textwidth]{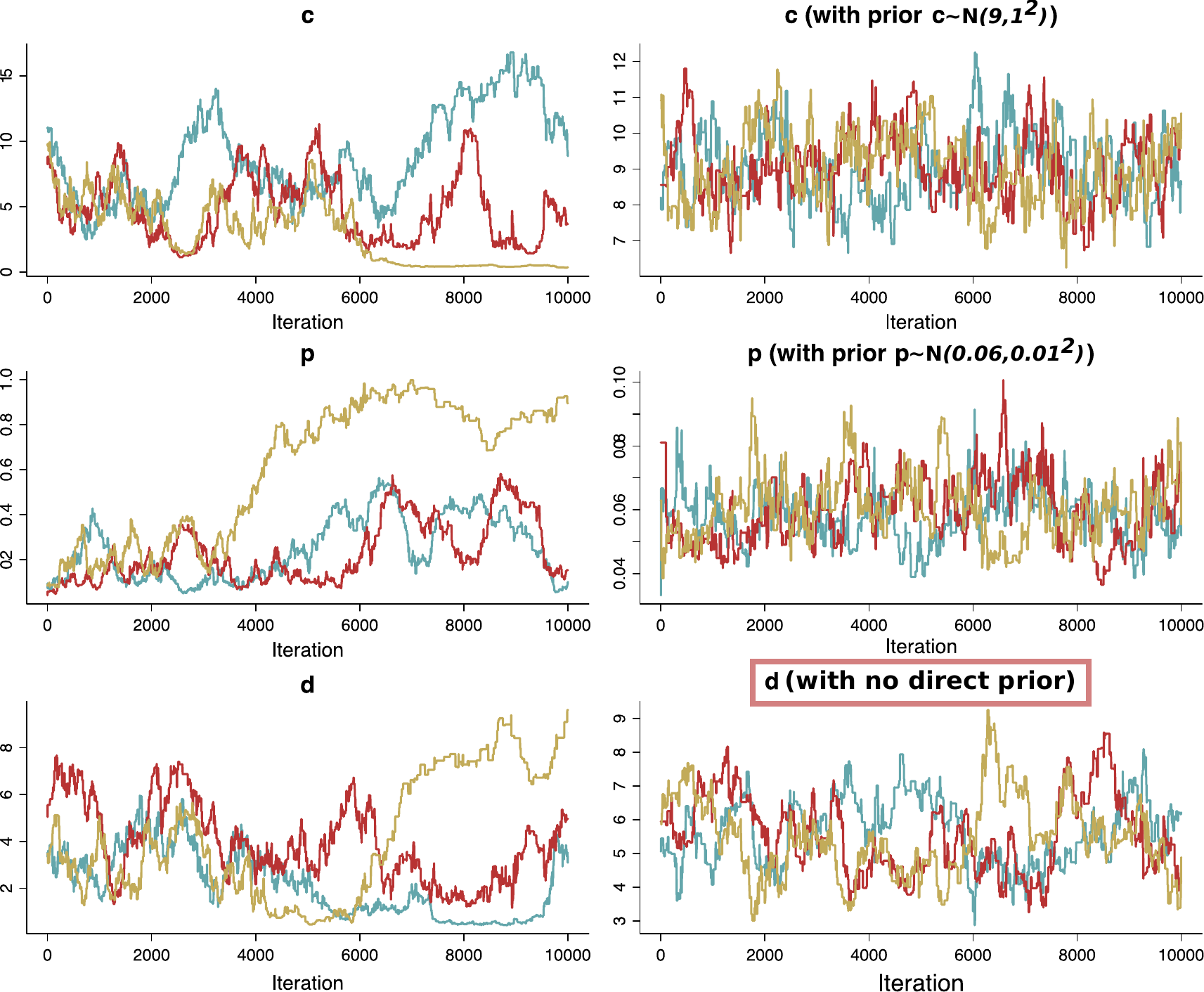}
 \caption{Resulting chains of running the Metropolis-Hastings algorithm with improper uniform priors and with informative priors for the SIS model.}
 \label{fig:SIS chains}
 \end{figure}
 
As identified by Alarid-Escudero \textsc{et al.} \cite{alarid2018nonidentifiability}, the main problem in models of this type is the lack of identifiability in the parameter space, given the information available. One solution is to obtain additional data or an additional target for which data may be available. The approach adopted here is to add additional information about the parameters in the form of prior information. This prior information is specified by using information from other sources (e.g.~from expert knowledge or previous experience of these parameters). Note that while the term `prior' may in ordinary usage be associated with information available before collecting data (i.e.~targets), in the Bayesian paradigm it is the fact that information other than contained in the targets may be incorporated that is important. The process of obtaining prior information is called prior elicitation and is a substantive area in its own right which is beyond the scope of this article. For interested readers, other literature such as a recent review\cite{mikkola2024prior}, work on the SHELF framework \cite{SHELFWebsite}, O’Hagan\cite{ohagan2019expert} and earlier articles such as Kadane and Wolfson\cite{kadane1998experiences} explore this area in more depth.

Consider a measles outbreak in a school setting. We aim to estimate the transmission rate ($\beta$) and recovery rate ($\gamma$) using an SIS model. Due to the limited data available from the school, the model may be susceptible to non-identifiability. To address the non-identifiability, we can incorporate prior information to constrain the parameter space. For instance for transmission rate ($\beta$) one could use expert opinion: based on previous measles outbreaks, experts might estimate that the basic reproduction number ($R_0$) is around 2.5. Since $R_0$ = $\beta$/$\gamma$, we can use this information to inform the prior distribution for $\beta$, assuming a reasonable recovery rate. Alternatively, one could look at the contact rate instead of the transmission rate itself. If we have data on the average number of contacts per student per day, we can use this to inform the prior for $\beta$, as $\beta = c \cdot p$ (where $c$ is the contact rate and $p$ is the transmission probability).

For recovery rate ($\gamma$) biological knowledge could be used. The average duration of measles infection is known, which can be used to set a prior distribution for $\gamma$ (e.g., a Beta distribution with appropriate parameters). By incorporating these prior distributions, we can help to narrow down the possible values for $\beta$ and $\gamma$, reducing the impact of non-identifiability. Sensitivity analysis can be performed to assess how the posterior distributions for $\beta$ and $\gamma$ are affected by different prior specifications.

Suppose we have a prior distribution for $\beta$ based on expert opinion and a prior distribution for $\gamma$ based on biological knowledge. By combining these priors with the likelihood function derived from the observed measles cases, we can obtain posterior distributions for $\beta$ and $\gamma$. If the posterior distribution for $\beta$ is relatively narrow, it suggests that the prior information has helped to resolve the non-identifiability issue.

One could start by using vaguely-informative uniform priors for each parameter. However, here we demonstrate how imposing an informative prior on some of the parameters improves identifiability and affects the convergence of the chains for other parameters implicitly. We impose normal priors for two parameters $c\sim N(9,1)$ and $p\sim N(0.06,0.01)$ effectively imposing a prior on $\beta$ indirectly (See right-hand side of \figref{fig:SIS chains} for results). The chains were compared using Gelman and Rubin's convergence diagnostic\cite{gelman1992inference}. The Gelman-Rubin diagnostic (or 
$\hat{R}$) assesses the convergence of multiple MCMC chains by comparing between-chain and within-chain variances. If $\hat{R}$ is close to 1, it indicates convergence, meaning the chains are sampling from the same posterior distribution. Values significantly greater than 1 suggest non-convergence, requiring more iterations or adjustments. The chains without an informative additional prior imposed scored a point estimate of 2.58 with an upper confidence interval of 5.05 and the chains with normal priors scored a 1.05 point estimate with upper C.I. of 1.15. 

\begin{figure}[h]
\begin{center}
\includegraphics[width=.99\textwidth]{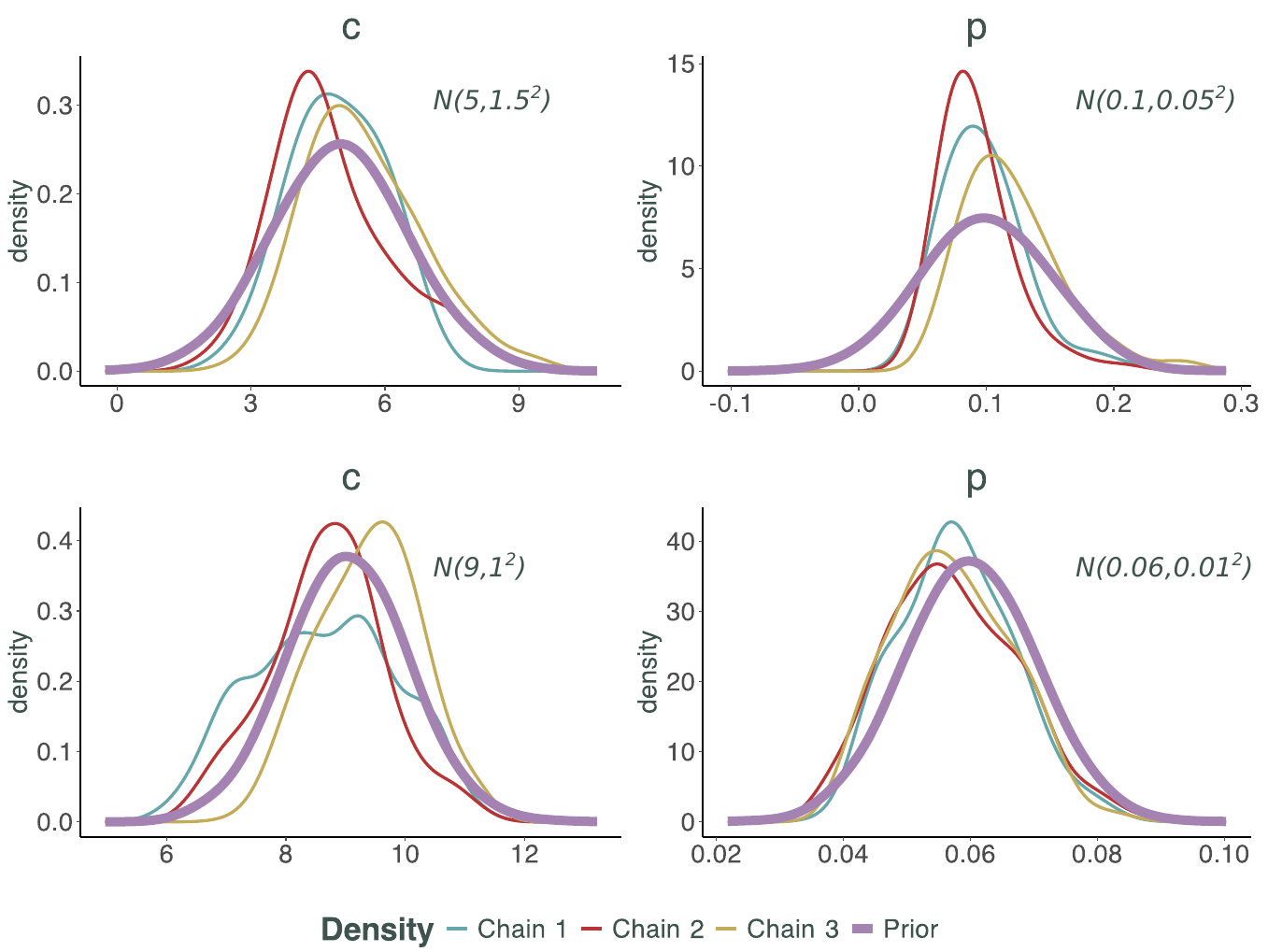}
\end{center}
\caption{Prior to posterior changes in distributions of 3 chains for selected parameters ($c$ and $p$) for which priors were specified directly in the SIS model calibration. All priors were normal distributions and means and standard deviations are specified on the plot. The bottom row of this figure corresponds to the right-hand panel of \figref{fig:SIS chains}.}
\label{fig:SIS_prior_to_posterior}
\end{figure}

While useful, $\hat{R}$ should be complemented by visual diagnostics like trace plots to ensure a thorough evaluation of convergence. In addition to trace plots presented in \figref{fig:SIS chains}, we also present a plot demonstrating the change from prior to posterior distributions for additional visual diagnostic. The posterior distributions for the parameters were very close to the imposed priors (\figref{fig:SIS_prior_to_posterior}), highlighting the fact that there is little or no information in the data itself about these parameters. 

\subsection{Prior sensitivities}

We conducted sensitivity analyses to assess how prior distributions influence posterior inferences. For the SIS model, we systematically varied the priors for $c$ (contact rate) and $p$ (transition probability) across a spectrum from weakly informative to strongly informative distributions. The results, as shown in \figref{fig:SIS_prior_to_posterior}, reveal a strong dependence of posterior estimates on the prior specifications. This indicates that the available data and model structure provide insufficient information to constrain parameter estimates, leading to wide credible intervals. These findings underscore the critical importance of carefully selecting priors, especially in scenarios with limited data or high non-identifiability, to ensure meaningful and robust inferences.

In summary, in this section, we have illustrated how the problem of non-identifiability can arise even in a relatively simple model, and shown how the use of additional information solves this problem. It is after initial runs that the modeller will detect that non-identifiability is occurring, and by carefully considering what information to add regarding other parameters, identifiability may be achieved. In practice, such additional information may come from, for example, monitoring a number of individual cases of the disease to better quantify the time to recovery. By carefully incorporating prior information, we can enhance the reliability and interpretability of disease models, even in the presence of non-identifiability. This case study demonstrates how prior knowledge can be used to guide parameter estimation and improve our understanding of disease dynamics.

\section{Example: HPV model}\label{sec:example}

The second model that we use in this paper to illustrate the challenges of non-identifiability in model calibration is an individual-based model, also called an agent-based model. An individual-based model simulates the characteristics of each individual and keeps track of each individual's history, recording all information relevant to the disease in question. The advantages of agent-based models are that they can incorporate heterogeneity (through each agent), they can be physically realistic, and they allow for interventions that impact individuals differently. They are commonly used when modelling dynamic aspects of infectious diseases as described for example in Hunter \textsc{et al.}\cite{hunter2017taxonomy}. The challenges associated with these models are that they take more computer time to run, they can be difficult to build, and of interest to this example, they can be harder to calibrate.

\begin{figure}[h]
\begin{center}
\includegraphics[width=.65\textwidth]{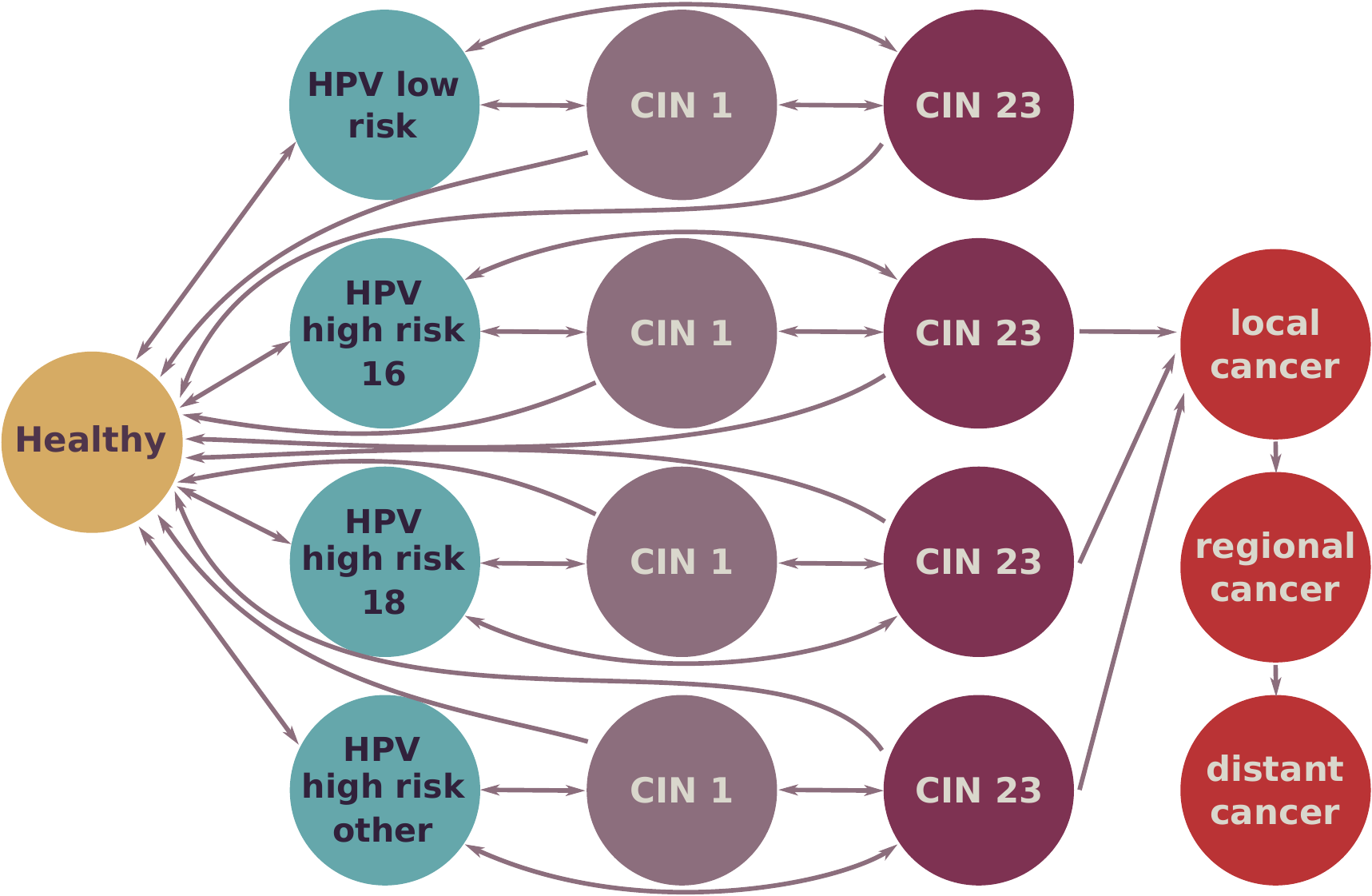}
\end{center}
\caption{Diagram of the HPV model modified from Kim \textsc{et al.} \cite{kim2007multiparameter}. The arrows represent transitions between health states (formulated in the model in terms of probabilities).}
\label{fig:diagram_HPV}
\end{figure}

The example we use here is based on the human papillomavirus (HPV) which causes cervical cancer. We follow the model developed by Kim \textsc{et al.} \cite{kim2007multiparameter} which considers transitions between Healthy state, HPV infection, pre-malignant changes called cervical intra-epithelial neoplasia (CIN) 1, CIN 2 and CIN 3 (which are different levels of severity) and, finally, cancer (see \figref{fig:diagram_HPV} for details). The HPV model incorporates a dynamic structure where individuals can move between different health states. Individuals can progress from a healthy state to HPV infection, and subsequently through various stages of cervical intraepithelial neoplasia (CIN) until reaching a cancer state. Importantly, the model also allows individuals to recover from HPV infection or regress from advanced CIN stages. Additionally, the model's parameters vary by strain (i.e.~transition probabilities to CIN states depend on the strain of HPV infection), and therefore the number of parameters for calibration is high. We have deliberately selected this HPV model for this paper because it is very complex and is likely to exhibit many of the challenges that were discussed in \Secref{Limitations} and \Secref{Non-Identifiability}.

Individual-based models, such as the HPV model presented here, offer a powerful framework for understanding the complex dynamics of infectious and other diseases\cite{rutter2009bayesian,perez2009agent,hunter2018open}. By simulating the behaviour of individual agents, these models can capture heterogeneity, non-linearity, and emergent phenomena. However, the complexity of these models, often coupled with limited and population-level data, can lead to challenges in model calibration, including non-identifiability. This makes it difficult to uniquely estimate model parameters, hindering our ability to draw definitive conclusions about the underlying biological processes. Addressing these challenges is crucial for obtaining reliable model predictions and informing effective public health interventions.

\subsection{Calibration using Metropolis-Hastings}\label{calibrating:M-H}

In this section, we present some results of applying the Metropolis-Hastings algorithm (Appendix \algref{alg:M-H}) to the HPV model calibration. Despite the advantages of the algorithm, these results illustrate some of the problems associated with calibrating complex models using the Metropolis-Hastings algorithm described in \Secref{Limitations} and \Secref{Non-Identifiability}.

\begin{figure}
\includegraphics[width=.99\textwidth]{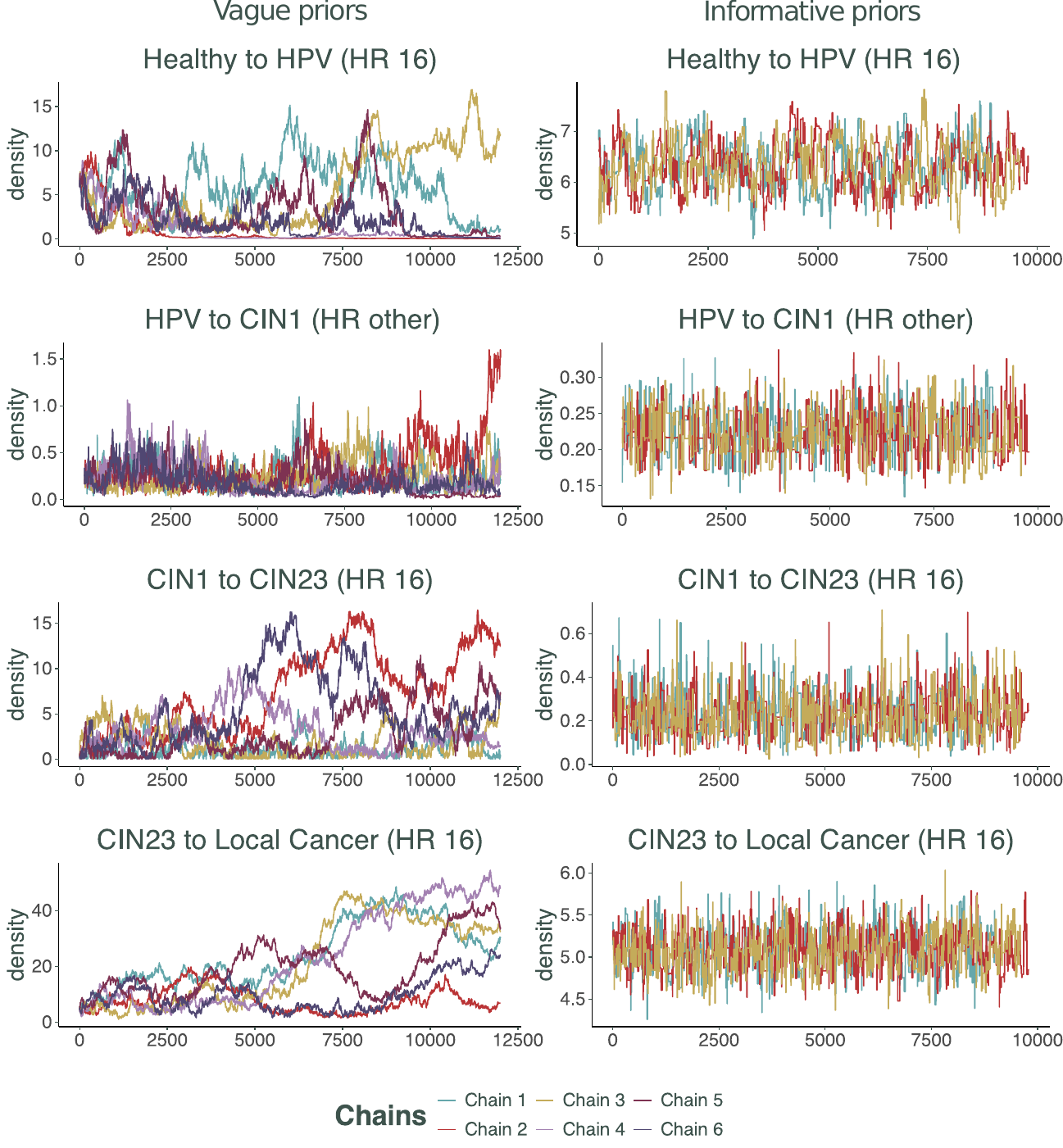}
\caption{Results of running Appendix \algref{alg:M-H} for 120,000 steps with \emph{improper uniform priors} (left-hand side) and 90,000 steps with \emph{Gamma priors} (right-hand side) for the HPV model.}
\label{fig:HPV_no_prior_vs_prior}
\end{figure}

This model has the same non-identifiability problem as the simple SIS model in \Secref{sec:example:SIS}. As was illustrated in \figref{fig:A_B_diagram} of \Secref{calib:M-H}, when you have two competing transition probabilities, successfully calibrating both can be impossible if the data available for calibration is population-averaged. In our case, the only data available were the distributions of population-level observations. This is a common problem of incompatibility of population-level data (the most commonly collected and available) with agent-based models. Attempting to calibrate this model with improper uniform priors will result in chains not converging as demonstrated by the left-hand side of \figref{fig:HPV_no_prior_vs_prior}. By only relying on visual diagnostics at this stage one could definitively conclude that those chains are not converging, indicating a potential problem with non-identifiability in the model. We demonstrate here how imposing even a reasonably vague prior on all parameters leads to great convergence of the chains produced by the algorithm.

To address the problem of non-identifiability of the parameters as described in\Secref{Non-Identifiability}, we introduce (vaguely-)informative priors. There are many ways to incorporate any information available at a given time into priors. If the priors for each parameter are assumed to be independent, then the prior $\pi(\boldsymbol{x})$ in Appendix \algref{alg:M-H} is just a product of one-dimensional priors $\pi(\boldsymbol{x})=\prod_{i=1}^{n}\pi_i(x_i),$ where each $\pi_i$ is a prior distribution for $i$'s parameter. The choice of priors represents our beliefs about parameters without taking new evidence into account.  Gamma distributions were chosen to be fitted to the parameters because of their $[0,\infty)$ support and their shape. In general, Beta priors would be a standard choice for probabilities due to their $[0,1]$ support. However, this model was designed such that the baseline probabilities of transitioning were hard-coded and multipliers to those baseline probabilities were controlled by the inputs to the model, leading to this slightly unintuitive choice of prior distribution family.  Priors on the model's parameters were imposed to limit the algorithm's search to (what we know to be) a plausible region of the space. The uniqueness of such a region can depend on many factors. Possible sources of information from which one could construct priors also include expert opinions and alternative data sets. 

We specify informative prior distributions for transition rates in our HPV model using the study by Jackson et. al \cite{jackson2015calibration}. This paper demonstrated the efficacy of using Bayesian methods to combine diverse evidence and propagate uncertainties in model calibration. Building on their results, we used Gamma distributions for prior specifications consistent with the identified intervals, considering their suitability for non-negative values and their flexibility in capturing various levels of uncertainty. By carefully selecting Gamma parameters based on expert knowledge and available data, we were able to incorporate valuable prior information into our model, enhancing the reliability of our parameter estimates and improving the overall robustness of our HPV model calibration. As highlighted above, the prior distribution can be thought of as ‘other information’ about the parameters than that provided by the likelihood. In this case, structural priors can be placed upon the parameter space by observing that large sections are not compatible with plausible data values. Plausible values had also to be chosen for the backward transitions. 

Additionally, instead of updating all parameters at once, on each step we randomly chose a subset of the parameter space of cardinality $s=7$ to be updated. This number was chosen on a trial-and-error basis and was constant throughout the implementation of the algorithm. The 7 parameters to be updated were chosen randomly at each step. That increased the acceptance rate of the algorithm and improved the mixing of the chain produced by Appendix \algref{alg:M-H}.

\begin{figure}[!tp]
\begin{center}
\includegraphics[width=.99\textwidth]{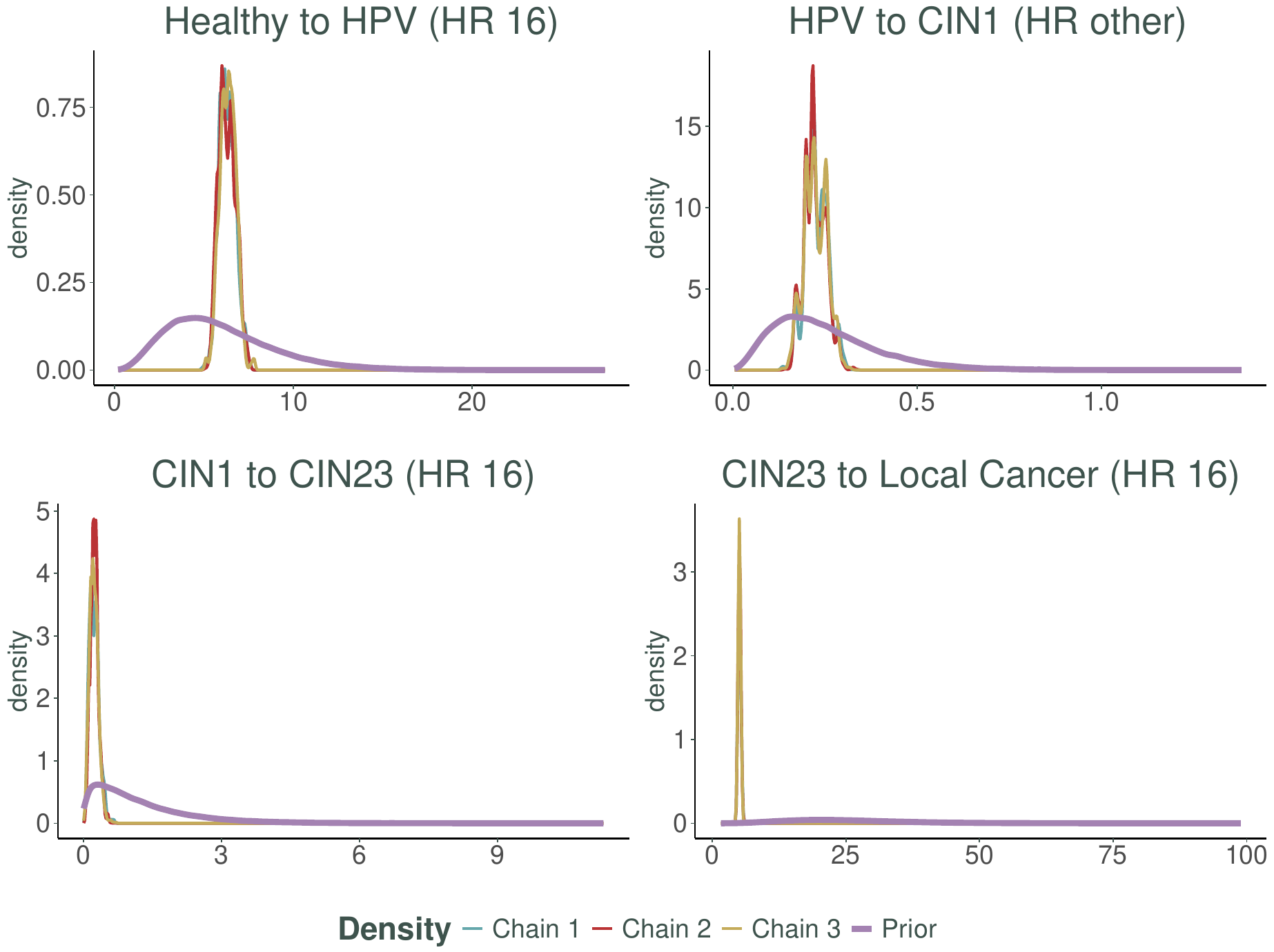}
\end{center}
\caption{Prior to posterior changes in distributions of 3 chains for selected 4 parameters for the HPV model. All priors were Gamma distributions. Posterior distribution plots for all parameters can be found in \appfigref{fig:result_desn_gamma_priors}.}
\label{fig:proposals_with_priors}
\end{figure}

The posterior distribution of the chains of the Metropolis-Hastings algorithm along with the corresponding priors for four selected parameters can be found in \figref{fig:proposals_with_priors}. Although here we focus on four of the most interesting parameters, the resulting posterior densities for all 26 parameters are presented in \appfigref{fig:result_desn_gamma_priors}. It is clear from \figref{fig:proposals_with_priors} that all the posterior distributions are distinct from the priors and in some cases, despite the somewhat vague priors, very distinct. This highlights again the point made in \Secref{sec:example:SIS} and \figref{fig:SIS chains}. In a complex model like this, priors of interacting parameters could have a knock-on effect. Even though there was not enough information in the data and the model itself to identify good parameter values, and the prior on this parameter was vague, there was enough information in other parameter's priors to have a knock-on effect and for the posterior distribution to have a pronounced peak, especially for CIN23 to Local Cancer (HR 16). 

It is also visible from the density plots in \figref{fig:proposals_with_priors} (and even more so in \appfigref{fig:result_desn_gamma_priors}) that some multi-modality exists in the posterior distributions. All three chains recognise the multi-modality and sampling from the same distribution. This multi-modality was not picked up by the original chain nor was it present in the Gamma priors. This highlights the importance of prior elicitation in complex models. The balance between informing some parameters from the data or informing them through the priors could be drastically shifted towards the priors in a situation of non-identifiability\cite{alarid2018nonidentifiability} or just general lack of information in the available data.

As a small aside we would like to make a point here that despite the complications associated with direct application of the Metropolis-Hastings algorithm (with improper uniform priors) to the HPV model calibration, compared to the original method of Kim \textsc{et al.} \cite{kim2007multiparameter} the goodness-of-fit (GOF) measure (see Appendix \ref{app:HPV} for details) was consistently better for all chains (low scores being better), staying near 20, fluctuating up and down around that value. Despite the problems, this was true for all approaches, even the ones that we have chosen not to present here. If we compare this to the simple Monte Carlo approach of Kim \textsc{et al.} \cite{kim2007multiparameter}, the best fitting parameter values had GOF values around 122 for the Irish data (with minimal value at 72 and maximal value at 185 in a sample of 133 best fitting values). The corresponding p-values are 93.58\% and $\ll 0.01\%$ for our approach (GOF=20) and Kim \textsc{et al.} \cite{kim2007multiparameter} approach (GOF=122) respectively. However, the fit of our approach appears to be almost `too good' relative to $\chi^{2}_{31}$. That can be explained by the correlation in the targets, so if one is fitting well, a few others are also fitting well. If we remove the assumption of independence, that measure of GOF would have a different distribution (depending on the correlations), but the mean of that distribution will be shifted towards zero compared to $E(\chi^2_{k=31})=31$. This will make the value of GOF=20 more realistic. 

By incorporating informative priors, we were able to effectively address the challenges of non-identifiability and improve the precision of our parameter estimates. We used Gamma distributions to specify priors for transition-rate multipliers, leveraging expert knowledge and available data. These priors helped to constrain the parameter space and prevent the algorithm from exploring implausible regions. As a result, the posterior distributions were more focused and informative, leading to more reliable model predictions and inferences for decision-makers.


\section{Discussion}\label{sec:discussion}

The Bayesian approach to model calibration is becoming more popular\cite{menzies2017bayesian} and more research is focusing on MCMC techniques in particular. Several calibration approaches become more difficult to apply in the context of stochastic models, such as stochastic microsimulations or agent-based models\cite{menzies2017bayesian}. Given the advantages of a Bayesian stochastic perspective, and in particular, here an MCMC approach, the fact that formal convergence is not always easily obtained without substantial additional work requires much attention. Such work includes better data collection or data representation in terms of priors as well as tuning the MCMC to the problem. 

A key message of this work is that good prior elicitation in complex models is essential for a good fit of the model's parameters. While here we have examined as an outcome the improvement in GOF, this must be considered in light of the expertise required to apply these methods to model calibration. Care is needed with starting values and proposal distributions to ensure that the chains have converged and that the parameter space is properly explored.

One of the main observations in this work was that it is possible to explicitly incorporate additional information into the inference in the form of priors during the calibration process and that this had a significant effect on the algorithm's performance. This has a big impact on the issue of non-identifiability. Once the additional information was included (here through the revision of the prior), the algorithm was not impaired by the presence of some local multi-modality and all three chains sampled from the same distribution. This led to a desirable effect of convergence that was coupled with the improved acceptance rate of the algorithm and a visible lack of autocorrelation in the sample.

The fact that the application of the Metropolis-Hastings algorithm yielded results with a lower GOF than the simple Monte Carlo approach is not surprising, given that the Metropolis-Hastings algorithm is designed to lead the sampler to a high posterior probability area, whether it is a global or a local maximum. Additionally, this technique provides probabilistic estimates of the parameters of interest and reflects their joint uncertainty. Usually, the acceptance criterion is applied after the fact, which leads to inefficiency.

In a population-level model of Hepatitis C infection, a similar process of calibration was performed \cite{Semochkina2018Bayesian}. This case had none of the problems that we had with the HPV model calibration. This was mostly because good priors were provided by an expert. Furthermore, because the Hepatitis C model did not have any backward transitions, there were fewer problems with identifiability, compared to the HPV model. Another possible advantage of that model was its relative simplicity, making it difficult to over-parameterise, compared to the individual-based models. This allowed us to try multiple strategies, before selecting the best one. We hypothesise that the relatively simpler structure of the Hepatitis C model, compared with the HPV model, yields a smoother probabilistic link between the model parameters and outputs. Of course, model choice is very influential in model calibration and different model types might yield different calibration results.

\section*{Funding}
This work was supported by the Science Foundation Ireland [grant number 12/IA/1683] and the Health Research Board [grant number RL13/04]. 

\section*{Acknowledgements}
We would like to thank MACSI (University of Limerick) and the Health Research Board who have provided the funding for this research. We are also grateful to Jane Kim for sharing the code from their paper\cite{kim2007multiparameter}. Daria Semochkina would also like to thank Paul A. Smith (for his advice on article preparation and many helpful discussions) and John G. Donohue (for tirelessly proofreading this article countless times). We would also like to thank the Editor-in-chief and anonymous reviewers for their helpful contribution to shaping this paper.


\bibliographystyle{unsrtnat}
\bibliography{Bib}


\appendix 
\section{Metropolis-Hastings algorithm}\label{app:M-H}
The central idea of \algref{alg:M-H} is to locally update, i.e.~use the accepted parameter value $\boldsymbol{x}_{(t)}$ at time step $t$ to generate the next proposed value. This makes it easier to think of a suitable proposal step (that is, of course, conditional), albeit with the disadvantage of producing a Markov chain (as opposed to an independent realisation). This Markov chain converges to the posterior distribution of interest after enough steps. The sample values are taken from the Markov chain after convergence. The basic steps are presented in Algorithm \ref{alg:M-H}.

\vspace{0.5cm}
\begin{algorithm}[h]
\caption{Metropolis-Hastings algorithm}
Starting from $\boldsymbol{x}_{(0)}=(x_{(0)}^1,x_{(0)}^2,...,x_{(0)}^n),$ for $t=1$ to $N$ do
\begin{algorithmic}[1]
\State Draw $\boldsymbol{x}\sim q(\cdot|\boldsymbol{x}_{(t-1)})$
\State Compute
\begin{equation}\label{eq:Met-Has}
p(\boldsymbol{x}|\boldsymbol{x}_{(t-1)})=\min\left\{1,\frac{f(\boldsymbol{x})q(\boldsymbol{x}_{(t-1)}|\boldsymbol{x})}{f(\boldsymbol{x}_{(t-1)})q(\boldsymbol{x}|\boldsymbol{x}_{(t-1)})}\right\}
\end{equation}
\State With probability $p(\boldsymbol{x}|\boldsymbol{x}_{(t-1)})$ accept the step and set $\boldsymbol{x}_{(t)}=\boldsymbol{x}$, otherwise reject the step and let $\boldsymbol{x}_{(t)}=\boldsymbol{x}_{(t-1)}$
\end{algorithmic}
End
\label{alg:M-H}
\end{algorithm}
\vspace{0.5cm}
Here $q$ is the \textbf{proposal distribution}, which is often difficult to choose. The probability of acceptance in \eqref{eq:Met-Has} does not depend on the normalising constants for either the proposal distribution $q$ or the distribution of interest $f$.

\section{HPV model}\label{app:HPV}
\renewcommand{\figurename}{Appendix Figure}
\renewcommand{\tablename}{Appendix Table}
\renewcommand\thefigure{\arabic{figure}}
\setcounter{figure}{0}
\renewcommand\thetable{\arabic{table}}
\setcounter{table}{0}

Here, we present further details about the HPV model, described in \Secref{sec:example}.
We assume that all of the targets have normal distributions $f_k\sim N(d_k,\sigma_k),$ centred at the observed data $d_k$ and with variances $\sigma_k$ that were calculated from the data. We assume all targets to be independent. Because of that assumption, our likelihood is just a product of individual likelihoods 
$$f(\boldsymbol{x})=\prod_{k=1}^K f_k(x_k)=\prod_{k=1}^K \frac{1}{\sigma_k\sqrt{2\pi}} e^{-\frac{1}{2}\left(\frac{x_k-d_k}{\sigma_k}\right)^2},$$
where $K=31$ is the number of calibration targets and $\boldsymbol{x}=(x_1,\ldots,x_K)$ are the model predictions. This product of likelihoods is scaled by each distribution probability density function at the mode and transformed onto a log scale to create a simplified goodness of fit (GOF) measure in the following way
\begin{equation*}
GOF_{total} = -2\sum_{k=1}^K\ln\left[\frac{f_k(x_k)}{f_k(d_k)}\right].
\end{equation*}
The details of the means and standard deviations can be found in \apptabref{tab:targ}. If we simplify the GOF measure, we can see that it is $\chi^2$ with 31 degrees of freedom (assuming independence).

\begin{equation}
\begin{split}
\nonumber
GOF_{total} &= -2\sum_{k=1}^{31}\ln\left[\frac{f_k(x_k)}{f_k(d_k)}\right]\\
&=-2\sum_{k=1}^{31}\ln\left(
\frac{
\frac{1}{\sqrt{2\pi\sigma_k^2}}\cdot e^{-\frac{(x_k-d_k)^2}{2\sigma_k^2}}
}
{
\frac{1}{\sqrt{2\pi\sigma_k^2}}\cdot e^{-\frac{(d_k-d_k)^2}{2\sigma_k^2}}
}
\right)\\
&=\sum_{k=1}^{31}\frac{(x_k-d_k)^2}{\sigma_k^2}=\sum_{k=1}^{31}\left(\frac{x_k-d_k}{\sigma_k}\right)^2\sim\chi^2_{k=31}.
\end{split}
\end{equation}

\begin{center}
\begin{table}[h]
\begin{scriptsize}
\centering
\begin{tabular}{p{8cm} rr}
\hline
\\
\normalsize{Calibration Target} & \normalsize{$d_k$} & \normalsize{$(\sigma_k)$}\\
\\
\hline
\\
\bf{Duration of HPV infection}\\\qquad \bf{(months)}\\
\\
\quad Low risk, $<$30 years & 9.775 & (0.599)\\
\quad High risk (other), $<$30 years & 11.720 & (0.638)\\
\quad High risk 16, $<$30 years & 12.905 & (1.334)\\
\quad High risk 18, $<$30 years & 10.015 & (1.620)\\
\quad Low risk, $>$30 years & 9.910 & (0.378)\\
\quad High risk (other), $>$30 years & 11.300 & (0.622)\\
\quad High risk 16, $>$30 years & 10.875 & (1.099)\\
\quad High risk 18, $>$30 years & 10.960 & (1.857)\\
\\
\bf{Prevalence of high-risk HPV}\\\qquad \bf{infection}\\
\\
\quad 20-24 years & 0.398 & (0.107)\\
\quad 25-34 years & 0.224 & (0.069)\\
\quad 35-44 years & 0.108 & (0.044)\\
\quad 45-54 years & 0.077 & (0.036)\\
\quad 55-64 years & 0.060 & (0.027)\\
\\
\bf{Distribution of HPV types}\\\qquad  \bf{among women with CIN1}\\
\\
\quad High risk 16 & 0.2222 & (0.025)\\
\quad High risk (other) & 0.1025 & (0.022)\\
\\
\bf{Distribution of HPV types}\\\qquad  \bf{among women with CIN23}\\
\\
\quad High risk 16 & 0.6333 & (0.148)\\
\quad High risk 18 & 0.0761 & (0.010)\\
\quad High risk (other) & 0.0557 & (0.055)\\
\\
\bf{Distribution of HPV types}\\\qquad  \bf{among women with cancer}\\
\\
\quad High risk 16 & 0.6810 & (0.029)\\
\quad High risk 18 & 0.1067 & (0.010)\\
\\
\bf{Incidence rate of invasive cancer}\\\qquad  \bf{ (cases per 100,000 per year)}\\
\\
\quad 25-29 years & 6.34 & (2.67)\\
\quad 30-34 years & 14.67 & (3.60)\\
\quad 35-39 years & 20.13 & (4.58)\\
\quad 40-44 years & 23.83 & (3.74)\\
\quad 45-49 years & 21.83 & (4.28)\\
\quad 50-54 years & 18.29 & (5.06)\\
\quad 55-59 years & 18.25 & (4.34)\\
\quad 60-64 years & 13.47 & (3.83)\\
\quad 65-69 years & 15.59 & (4.08)\\
\quad 70-74 years & 14.46 & (4.07)\\
\quad 75-79 years & 16.38 & (6.74)\\
\\
\hline
\end{tabular}
\caption{Calibration targets with corresponding distribution parameters ($d_k$ is the mean and $\sigma_k$ is the standard deviation). The mean and variance for each parameter were elicited from the available data independently for lack of a better alternative.}
\label{tab:targ}
\end{scriptsize}
\end{table}
\end{center}

We opted not to use the calibration ranges adopted by Kim \textsc{et al.} \cite{kim2007multiparameter}. The Metropolis-Hastings algorithm leads the search in the direction of good-fitting values. Consequently, we do not need to worry about our computational budget to be able to cover the entire space. It is apparent that although some of the parameters are comparable with the results in Kim \textsc{et al.} \cite{kim2007multiparameter}, there is variation even between the two approaches presented in this paper. Transitions from HPV to CIN1 (HR 16); CIN1 to CIN23 (LR); CIN23 to CA (HR other) and CIN23 to CA (HR 18) have much higher values with the independent proposals strategy, compared to the strategy with gamma priors. Immune Degree (LR) has very different values in the two different strategies (this was assumed to be zero in the original paper). These results illustrate some of the concepts presented in \Secref{calib:M-H}, highlighting that different parameter combinations can give similar results.

\begin{figure}[!ht]
 \begin{center}
\includegraphics[width=.9\textwidth]{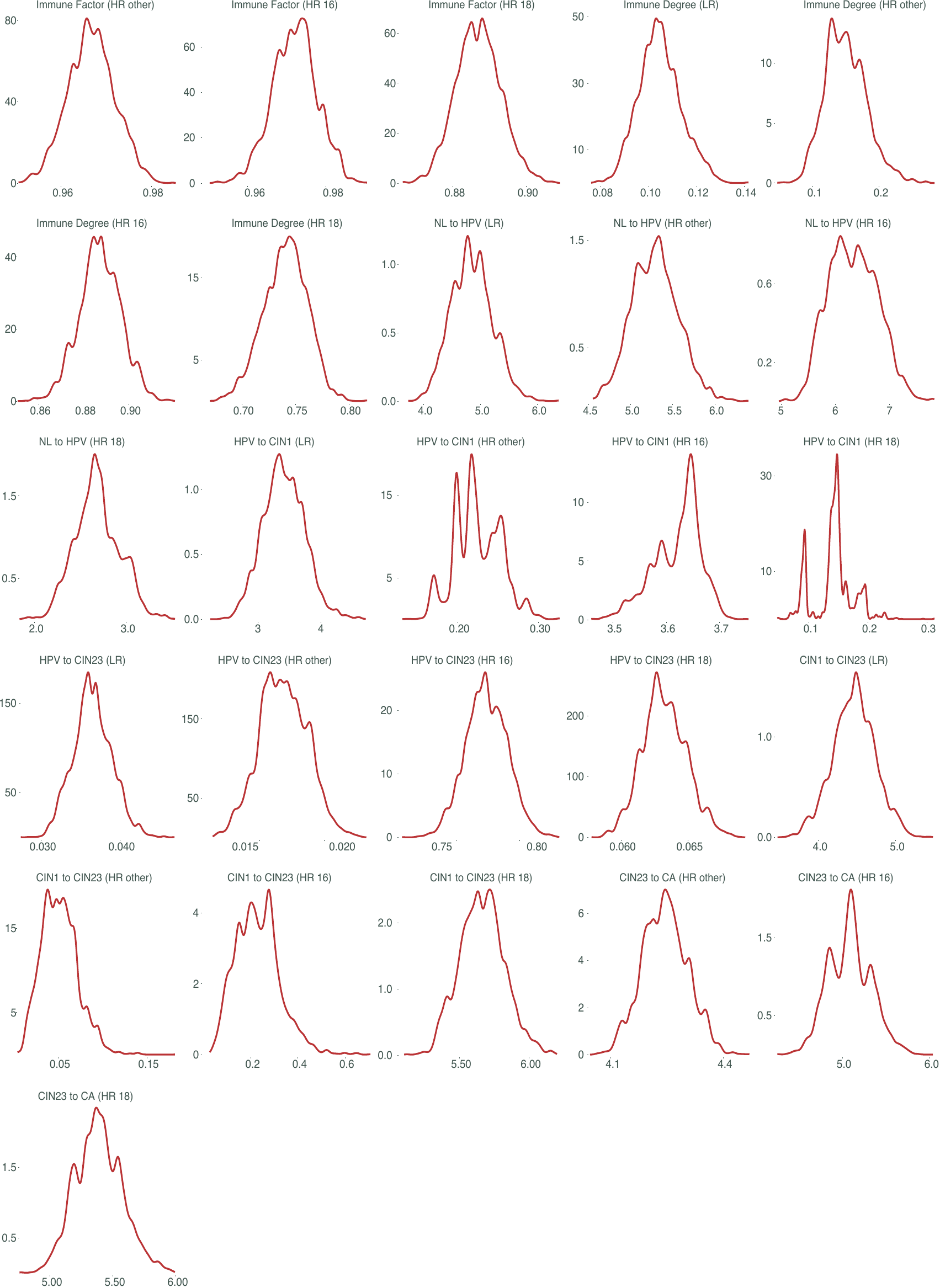}
 \end{center}
\caption{Posterior densities of all parameters with partial independent proposals, gamma priors and GOF$\leq25$. See \Secref{calibrating:M-H} for full details of the approach.}
\label{fig:result_desn_gamma_priors}
\end{figure}

Additionally, we present posterior distributions of all parameter sets with the corresponding GOF below 25 (see \appfigref{fig:result_desn_gamma_priors}).

\end{document}